\documentclass[nofootinbib,twocolumn,showpacs,preprintnumbers,pre,aps,superscriptaddress]{revtex4-2}

\usepackage[dvipdfmx]{graphicx}% Include figure files eps, pdf etc.

\usepackage{bm}% bold math
\usepackage{amsmath}
\usepackage{amssymb}
\usepackage{color}
\usepackage{newtxtext}
\usepackage{newtxmath}

\begin{document}

\title{Geometric formulation of state-dependent Langevin dynamics using scalar free energy}

\author{Kento Yasuda}\email{yasuda.kento@nihon-u.ac.jp}
\affiliation{Laboratory of Physics, College of Science and Technology, Nihon University, Funabashi, Chiba 274-8501, Japan}

\author{Zhongqiang Xiong}
\affiliation{Zhejiang Key Laboratory of Soft Matter Biomedical Materials, Wenzhou Institute, University of Chinese Academy of Sciences, 
Wenzhou, Zhejiang 325001, China}

\author{Zhanglin Hou}
\affiliation{Zhejiang Key Laboratory of Soft Matter Biomedical Materials, Wenzhou Institute, 
University of Chinese Academy of Sciences, 
Wenzhou, Zhejiang 325001, China}

\author{Kenta Ishimoto}
\affiliation{Department of Mathematics, Kyoto University, Kyoto 606-8502, Japan}

\author{Xinpeng Xu}
\affiliation{Department of Physics and MATEC Key Lab, Guangdong Technion - Israel Institute of Technology, Shantou, Guangdong 515063, China}
\affiliation{Technion - Israel Institute of Technology, Haifa 32000, Israel}

\author{Shigeyuki Komura}\email{komura@wiucas.ac.cn}
\affiliation{Zhejiang Key Laboratory of Soft Matter Biomedical Materials, Wenzhou Institute, University of Chinese Academy of Sciences, 
Wenzhou, Zhejiang 325001, China}

\begin{abstract}
Stochastic dynamics with state-dependent diffusion arise broadly in confined, anisotropic, and hydrodynamically coupled systems.
The conventional Langevin equation contains a spurious drift associated with multiplicative noise.
Since the restricted free energy is generally non-scalar, the covariance is not explicit.
Here, we reformulate state-dependent Langevin dynamics using a scalar free energy and a diffusion metric defined by the inverse 
diffusion tensor.
The conventional spurious drift is then expressed as a Christoffel contribution.
Although our formulation is mathematically equivalent to the conventional one through the relation between the restricted and 
scalar free energies, it makes coordinate covariance explicit and enables the phenomenological construction of thermodynamic 
potentials directly from system symmetries.
We demonstrate its consistency in representative examples of state-dependent diffusion arising from coordinate transformations, 
geometric confinement, and projection from curved to flat spaces.
\end{abstract}

\maketitle

%%%%%%%%%%%%%
\textit{Introduction}--
%%%%%%%%%%%%%
Stochastic dynamics with non-uniform or anisotropic state-dependent diffusionare widespread in soft matter, including confined 
Brownian particles, membrane inclusions, hydrodynamically coupled colloids, and molecular-conformation dynamics~ \cite{DoiBook,Komura2015,Yasuda2016,Naji2007,Ohta2020,Kobayashi2021,Alexandre2023NonGaussian,Best2010Coordinate,Sicard2021Position}.
In addition, apparent state-dependent diffusion can arise from coordinate transformations, such as those from Cartesian to polar coordinates, 
or from the geometry of the configuration space.
Such dynamics are commonly described by Langevin equations (LEs) and Fokker--Planck equations (FPEs) with multiplicative noise~\cite{RiskenBook}. 
In the presence of state-dependent diffusion, reproducing the canonical distribution requires an additional drift term associated with the diffusion tensor~\cite{Ermak1978Brownian,Morse2004Theory,Lau07,Volpe2016}. 
A particularly well-known formulation is the LE proposed by Lau and Lubensky (LL-LE), in which the drift correction is expressed 
by the gradient of the diffusion tensor, $\partial_jD^{ij}$~\cite{Lau07}. 
Their formulation has become a practical standard in Brownian dynamics and stochastic thermodynamics~\cite{Shiraishi2023}, 
and has been widely used in studies of confined diffusion, active matter, and 
stochastic hydrodynamics~\cite{Kuroiwa14,Itami2017,Cates2021Stochastic,Pigolotti2017,Sorkin2023,Frishman2020,Dechant2020,Avni2021}.

The thermodynamic behavior of stochastic systems is generally characterized by a free-energy functional, whose gradient determines 
the irreversible drift together with the mobility coefficients~\cite{DoiBook}.
Such dissipative dynamics can be systematically derived from Onsager's variational  
principle~\cite{Onsager31a,Yasuda24,Yasuda25,yasuda2026covariant}, 
where the free energy plays the role of a thermodynamic potential. 
In this context, free-energy functionals are naturally treated as scalar quantities reflecting the symmetries of the physical 
system~\cite{ChaikinBook,VitelliBook}. 
However, Uneyama and Ding et al.\ pointed out that the free energy appearing in the conventional 
LL-LE is a restricted free energy that generally transforms with a log-Jacobian term under point transformations, 
unlike the scalar free energy~\cite{Uneyama20,Ding2020}. 
This transformation law prevents the phenomenological construction of the free energy based on standard symmetry considerations,
as in the construction of a Landau free energy~\cite{ChaikinBook,VitelliBook}.
As a consequence, the covariance of the stochastic dynamics is not explicit when using the restricted free energy, and coordinate transformations require special care even in simple coordinate systems.

Although geometric approaches to LEs have been discussed by Graham and others~\cite{Graham77,Graham1985,Polettini2013,Diosi24}, their formulations did not explicitly address the role of thermodynamic potentials.
Nakamura proposed a scalar free energy by identifying the diffusion tensor as a metric on the configuration space~\cite{Nakamura24}; 
however, the corresponding LEs were not investigated.
In this Letter, by unifying Graham's equations with Nakamura's covariant thermodynamics, we reformulate state-dependent LEs in terms of a scalar free energy and a diffusion metric.
Within our framework, the conventional spurious drift is naturally rewritten as a Christoffel contribution associated with the 
diffusion tensor.

A significant advantage of the proposed LEs is that they enable the construction of a thermodynamic potential based on standard symmetry 
arguments.
The resulting covariant LE (C-LE) is mathematically equivalent to the conventional LL-LE
once the relation between the non-scalar and scalar free energies is taken into account.
However, in our formulation, the covariance of the dynamics is explicit because the C-LE is expressed in terms of a scalar thermodynamic potential.
Our formulation clarifies the geometric origin of the spurious drift and provides a consistent framework for coordinate transformations in
stochastic dynamics with state-dependent diffusion.
The theory is illustrated by Brownian motion in polar coordinates, on a spherical surface, and on a curved surface.

%%%%%%%%%%%%%%%%%%%%%
\textit{Restricted free energy}--
%%%%%%%%%%%%%%%%%%%%%
We consider a system in contact with a heat bath at temperature $T$.
The microscopic variables of the system, such as the positions and momenta of all particles, 
are denoted by $\mathbf {x}$. 
The Hamiltonian representing the energy of the microstate $\mathbf{x} $ is $H(\mathbf{x})$.
In irreversible thermodynamics, we are interested in several mesoscopic variables $\mathbf{q}$, 
such as the position of colloidal particles, rather than in all of $\mathbf{x}$. 
We assume that the mesoscopic variables $\mathbf {q} $ are uniquely determined when all of the 
microscopic variables $\mathbf{x}$ are given, i.e., $\mathbf{q} = \mathbf{q}(\mathbf x)$.

The statistical characterization of $\mathbf{q} $ can be given by its probability density function (PDF) $P(\mathbf q,t)$, 
which evolves with time $t$ and satisfies the normalization condition $\int d\mathbf{q}\, P(\mathbf{q},t)=1$.
In equilibrium, the PDF converges to the canonical distribution,  
$P_\mathrm{eq}(\mathbf{q})=N\exp\left[-F(\mathbf{q})/(k_\mathrm BT)\right]$,
where $N$ is a normalization constant, $k_\mathrm B$ is the Boltzmann constant, and $F(\mathbf{q})$ is the restricted 
free energy~\cite{DoiBook}
\begin{align}
F(\mathbf{q})=-k_\mathrm BT\ln \int d\mathbf x\, e^{-H(\mathbf x)/k_{\rm B}T}\delta(\mathbf{q}-\mathbf{q}(\mathbf{x})).
\label{restrictedfe}
\end{align}
Although the restricted free energy is formally defined by Eq.~(\ref{restrictedfe}),
it is generally not derived from this microscopic definition in practice.
Instead, it is usually constructed phenomenologically based on system symmetries~\cite{ChaikinBook,VitelliBook}.
For such a phenomenological construction, it is important to examine whether the free energy is invariant under a change of coordinates.

Let us therefore consider an invertible point transformation from $\mathbf{q}$ to $\mathbf{Q}$,
defined by $\mathbf{Q}=\mathbf{Q}(\mathbf{q})$, with inverse
$\mathbf{q}=\mathbf{q}(\mathbf{Q})$.
Under this transformation, the PDF transforms as a density rather than as a scalar; in general,
$P(\mathbf{q},t)\ne P(\mathbf{Q},t)$.
Consequently, the restricted free energy is not a scalar quantity either,
i.e., $F(\mathbf{q})\ne F(\mathbf{Q})$, but acquires a log-Jacobian contribution
under the coordinate transformation~\cite{Uneyama20,Ding2020,Nakamura24}.
This transformation law obstructs the phenomenological construction of the free energy
based on symmetry arguments.

%%%%%%%%%%%%%%%%%%
\textit{Diffusion tensor}--
%%%%%%%%%%%%%%%%%%
For phenomenological modeling, it is advantageous to introduce a scalar free energy instead of the non-scalar one.
To formulate such a scalar free energy, we first introduce the contravariant diffusion tensor $D^{ij}(\mathbf q)$, defined through the short-time covariance $\langle dq^i dq^j\rangle=2D^{ij}dt$, 
where $\langle\cdots\rangle$ denotes the noise average~\cite{RiskenBook}.
Hereafter, superscripts and subscripts denote contravariant and covariant components, respectively, 
and we use the Einstein summation convention for repeated indices. 
The diffusion tensor $D^{ij}$ is symmetric and positive definite~\cite{Onsager31a}, and is related to the mobility tensor 
$\mu^{ij}$ through the fluctuation-dissipation relation 
$D^{ij}=k_{\rm B}T\mu^{ij}$~\cite{DoiBook}.

Next, we introduce the Riemannian manifold $(M,D^{-1})$, where $M$ denotes the abstract configuration space
and $\mathbf{q}$ provides a local coordinate chart on $M$ representing the mesoscopic state.
The metric of this manifold is given by the inverse diffusion tensor
$D_{ij}$~\cite{Graham77,Graham1985,Polettini2013,Nakamura24},
i.e., $D^{ij}D_{jk}=\delta^i_k$, where $\delta^i_k$ is the Kronecker delta.
Hereafter, we refer to $D_{ij}$ as the diffusion metric to emphasize that it is determined by the diffusion tensor.
The diffusion metric characterizes how stochastic fluctuations vary from point to point and determines the local diffusive time scale $dt \sim dq^iD_{ij}dq^j$ associated with the displacement $d\mathbf q$. 
 This quadratic form is invariant under point transformations, which is a key property in making the covariance of a stochastic process explicit.

%%%%%%%%%%%%%%%%%%%
\textit{Scalar free energy}--
%%%%%%%%%%%%%%%%%%%
On the manifold $(M, D^{-1})$, we introduce the scalar probability density 
$\rho(\mathbf q,t)=P(\mathbf q,t)/\sqrt{\Delta(\mathbf q)}$, 
where $\Delta=\det D_{ij}$ is the determinant of the diffusion metric~\cite{Graham77,Graham1985,Polettini2013}.
The normalization condition for $\rho$ is $\int dV\, \rho=1$,
where $dV=\sqrt{\Delta}\,d\mathbf{q}$ is the invariant volume element 
(note that $\rho dV=Pd\mathbf q$).
Let us introduce the scalar free energy $\mathcal F$ via the reformulated canonical distribution~\cite{Nakamura24},
\begin{align}
\rho_\mathrm{eq}(\mathbf q)=N\exp\left(-\frac{\mathcal F(\mathbf q)}{k_\mathrm BT}\right),
\label{Eq:CD}
\end{align}
where both sides are scalars.

Importantly, unlike the restricted free energy $F$, the scalar free energy $\mathcal F$
can be constructed phenomenologically from tensorial quantities consistent with the
symmetries of the system~\cite{ChaikinBook,VitelliBook}.
For example, the scalar free energy of a free particle can be set to the coordinate-invariant
value $\mathcal F=0$, whereas $F$ cannot in general be chosen to be coordinate invariant.
This distinction is particularly important in curved configuration spaces, where no single
Cartesian coordinate chart is available.
Thus, $\mathcal F$ provides a natural basis for phenomenological modeling, as further
illustrated in later examples.

Once the scalar free energy $\mathcal F$ is specified phenomenologically,
the restricted free energy $F$ is obtained as~\cite{Nakamura24}
\begin{align}
F(\mathbf q)=\mathcal F(\mathbf q)-k_\mathrm BT\ln \sqrt{\Delta}.
\label{FreeEnergy}
\end{align}
This relation follows directly from Eq.~(\ref{Eq:CD}) and the original canonical distribution,
and shows that $F$ transforms with a log-Jacobian contribution.

%%%%%%%%%%%%%%%%%%%%%%
 \textit{Fokker-Planck equation (FPE)}--
 %%%%%%%%%%%%%%%%%%%%%%
The FPE determines the time evolution of the scalar probability density $\rho$ and can be written as~\cite{Graham77,RiskenBook} 
\begin{align}
\partial _t \rho =\nabla_i\left(\frac{D^{ij}\partial_j \mathcal F}{k_\mathrm BT}\rho+D^{ij}\partial_j\rho\right),
\label{Eq:FPeqrho}
\end{align}
where $\partial_t=\partial/\partial t$ and $\partial_i=\partial/\partial q^i$.
The covariant derivative is defined as 
$\nabla_i A^j=\partial_iA^j+\Gamma_{ik}^jA^k$, where 
$\Gamma_{ik}^j$ are the Christoffel symbols associated with the diffusion metric~\cite{Wald84,Graham77}. 
\begin{align}
\Gamma_{jk}^i=\frac{1}{2} D^{il}\left(\partial_j D_{kl}+\partial_k D_{jl}-\partial_l D_{jk}\right).
\end{align}
One can confirm that the canonical distribution in Eq.~(\ref{Eq:CD}) is a steady-state solution of Eq.~(\ref{Eq:FPeqrho}).
We also note that a constant rescaling of the diffusion tensor does not affect the Christoffel symbols, but only shifts the scalar 
free energy by an additive constant.
In the presence of a non-conservative force $f_j$, we can replace $\partial_j \mathcal F$ with 
$\partial_j \mathcal F-f_j$.

Substituting the scalar probability density $\rho(\mathbf q,t)=P(\mathbf q,t)/\sqrt{\Delta(\mathbf q)}$ into Eq.~(\ref{Eq:FPeqrho}) 
and using the relation $\partial_i (1/\sqrt{\Delta})=-\Gamma_{ij}^j /\sqrt{\Delta}$, we have 
\begin{align}
\partial _t P=\sqrt{\Delta}\nabla_i\left[\frac{1}{\sqrt{\Delta}}\left(\frac{D^{ij}\partial_j \mathcal F}{k_\mathrm BT}P-D^{ij}\Gamma_{jk}^kP+D^{ij}\partial_jP\right)\right].
\end{align}
Rewriting the covariant derivative, we obtain  
\begin{align}
\partial _t P=\partial_i\left(\frac{D^{ij}\partial_j \mathcal F}{k_\mathrm BT}P-D^{ij}\Gamma_{jk}^kP+D^{ij}\partial_jP\right).
\end{align}

Finally, using the metric compatibility condition $\nabla_k D^{ij}=\partial_kD^{ij}+\Gamma_{kl}^iD^{jl}+\Gamma_{kl}^jD^{il}=0$~\cite{Wald84,Graham77},
we arrive at the FPE for the non-scalar PDF $P$ as 
\begin{align}
\partial _t {P}=\partial_i\left[\left(\frac{D^{ij}\partial_j \mathcal F}{k_\mathrm BT}+\Gamma_{jk}^iD^{jk}\right) P+\partial_j (D^{ij} P)\right].
\label{Eq:FPeqP}
\end{align}
The corresponding canonical distribution $P_\mathrm{eq}(\mathbf q) = \sqrt{\Delta} \rho_\mathrm{eq}(\mathbf q) =
N \exp[-F(\mathbf q)/(k_\mathrm BT)]$ is the steady-state solution of Eq.~(\ref{Eq:FPeqP}).
In the above, the term $\Gamma_{jk}^iD^{jk}$ appears because the volume element varies over configuration space for 
state-dependent diffusion.

%%%%%%%%%%%%%%%%%%
\textit{Langevin equation (LE)}--
%%%%%%%%%%%%%%%%%%
Let us write down the corresponding covariant LE (C-LE) for the mesoscopic variables,
$\dot{q}^i=dq^i/dt$, that is consistent with the canonical distribution.
The C-LE is derived from the FPE~(\ref{Eq:FPeqP}) for $P(\mathbf q,t)$,
since this form provides a direct correspondence with the standard FPE--LE relation
and facilitates comparison with the LL formulation.

By following the standard procedure~\cite{RiskenBook}, the C-LE is derived as follows:
\begin{align}
\dot q^i=-\frac{D^{ij}\partial_j \mathcal F}{k_\mathrm BT}-\Gamma_{jk}^iD^{jk}
-2C B_{\alpha}^j\partial_j B_{\alpha}^i+\sqrt{2} [B_{\alpha}^i \bullet \xi_\alpha]_C,
\label{Eq:LangevinA}
\end{align}
which constitutes the main result of this Letter.
In the above, $B_\alpha^i$ is a noise amplitude satisfying $D^{ij}=B_{\alpha}^i B_{\alpha}^{j}$, and is defined up to an arbitrary
orthogonal transformation.
The Greek index $\alpha=1,2,\cdots$ labels components with respect to a local
orthonormal basis of the tangent space of the configuration space.
Since $\alpha$ labels orthonormal-frame components rather than coordinate components,
no distinction between covariant and contravariant Greek indices is required.
We consider homogeneous isotropic Gaussian white noise in the tangent space with 
$\langle\xi_\alpha(t)\xi_\beta(0)\rangle=\delta_{\alpha \beta}\delta(t)$.
Furthermore, the stochastic product is defined by $[A \bullet \xi]_C=[(1-C)A(\mathbf q(t))+CA(\mathbf q(t+dt))]\xi(t)$, where $\xi(t)dt$ represents the Wiener increment over the interval $[t,t+dt]$ and $0\le C \le 1$ is a stochastic convention parameter; 
$C=0$ and $C=1/2$ correspond to the It\^o and Stratonovich conventions, respectively.

For the It\^o convention ($C=0$), the C-LE~(\ref{Eq:LangevinA}) becomes
\begin{align}
\dot q^i=-\frac{D^{ij}\partial_j \mathcal F}{k_\mathrm BT}-\Gamma_{jk}^iD^{jk}
+\sqrt{2} [B_{\alpha}^i\bullet \xi_\alpha]_0,
\label{Eq:LangevinIto}
\end{align}
whereas for the Stratonovich convention ($C=1/2$), it becomes
\begin{align}
\dot q^i=-\frac{D^{ij}\partial_j \mathcal F}{k_\mathrm BT}-B_{\alpha}^j\nabla_j B_{\alpha}^i
+\sqrt{2}[B_{\alpha} ^i\bullet \xi_{\alpha}]_{1/2}.
\label{Eq:LangevinS}
\end{align}
Equation~(\ref{Eq:LangevinS}) can also be derived by applying a timescale decomposition to the underdamped LE 
with a standard potential energy~\cite{Cai24}.
Although Eq.~(\ref{Eq:LangevinA}) represents the same stochastic process for any choice of the stochastic convention parameter $C$,
its geometric structure is manifest in the Stratonovich form, Eq.~(\ref{Eq:LangevinS}).
Here, the stochastic increment obeys the standard chain rule under point transformations,
while the drift term is expressed geometrically through the covariant derivative,
$\nabla_j B_{\alpha}^i=\partial_j B_{\alpha}^i+\Gamma_{jk}^i B_{\alpha}^k$.

Using Eq.~(\ref{FreeEnergy}) to express the scalar free energy $\mathcal F$ in terms of the restricted free energy $F$
and substituting it into the C-LE~(\ref{Eq:LangevinA}), we recover the LL-LE~\cite{Lau07},
\begin{align}
\dot q^i=-\frac{D^{ij}\partial_j F}{k_\mathrm BT}+\partial_jD^{ij}-2C B_{\alpha}^j\partial_j B_{\alpha}^i
+\sqrt{2}[B_{\alpha}^i\bullet \xi_{\alpha}]_C.
\label{Eq:LangevinLau}
\end{align}
This LE is widely used in studies of state-dependent diffusion and stochastic thermodynamics~\cite{Kuroiwa14,Itami2017,Cates2021Stochastic,Pigolotti2017,Sorkin2023,Frishman2020,Dechant2020,Avni2021}.
However, because the restricted free energy $F$ in Eq.~(\ref{Eq:LangevinLau}) is generally not a scalar under point transformations~\cite{Uneyama20,Ding2020},
the covariance of the dynamics is not explicit in this representation.
When Eq.~(\ref{FreeEnergy}) is taken into account, the C-LE~(\ref{Eq:LangevinA}) and LL-LE~(\ref{Eq:LangevinLau})
describe the same stochastic dynamics but decompose the drift differently:
the spurious drift is $-\Gamma^i_{jk}D^{jk}$ in the C-LE and $\partial_jD^{ij}$ in the LL-LE.
This distinction in the drift decomposition persists even in a 1D configuration space.

\begin{figure}[t]
\begin{center}
\includegraphics[scale=0.4]{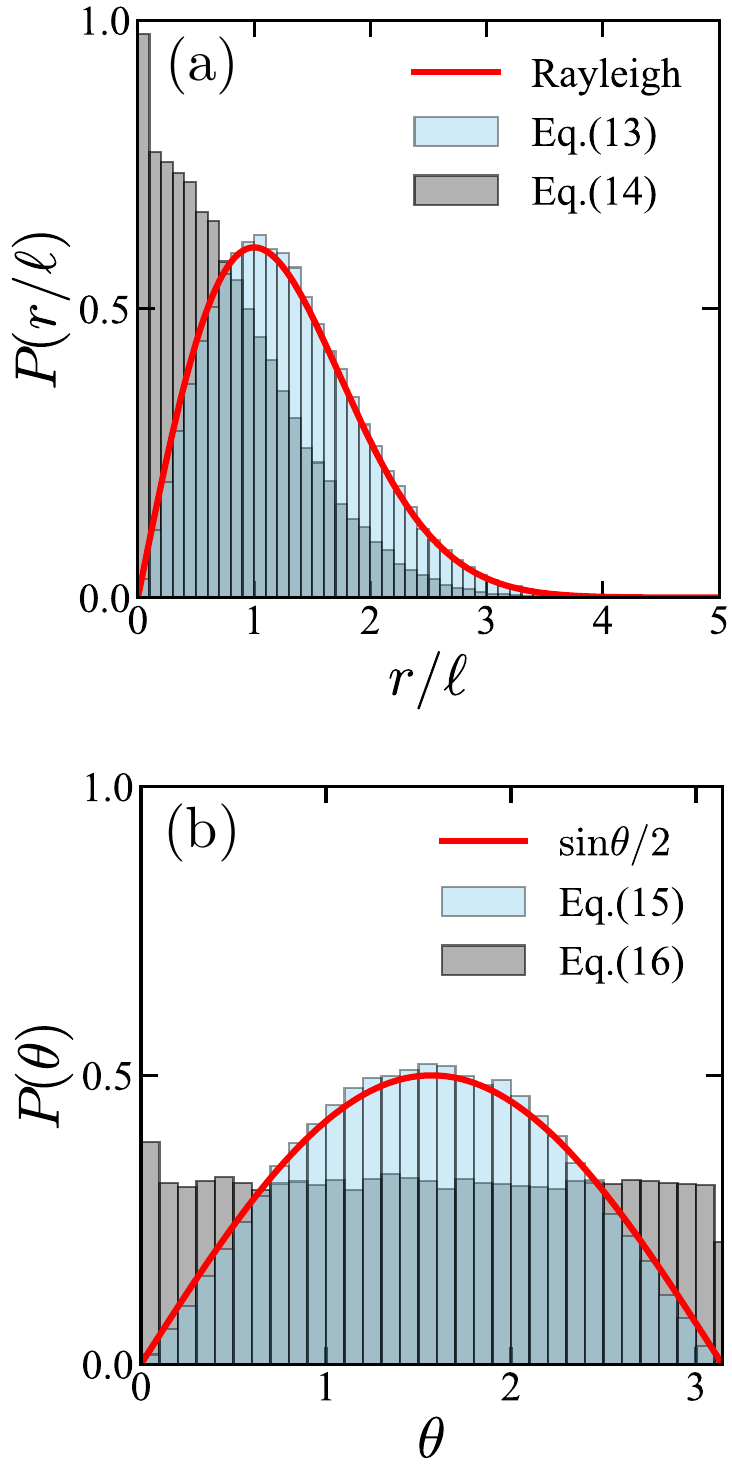}
\end{center}
\caption{
(a) PDF of the radius $P(r)$ in a 2D plane.
The blue and gray distributions are obtained from the correct Eq.~(\ref{Eq:r}) and the incorrect
Eq.~(\ref{Eq:r-Lau}), respectively.
The red solid line represents the expected Rayleigh distribution.
The length and time scales are $\ell=\sqrt{D/(\mu K)}$ and $\tau=1/(\mu K)$, respectively.
(b) PDF of the polar angle $P(\theta)$ on a spherical surface.
The blue and gray distributions are obtained from the correct Eq.~(\ref{Eq:theta}) and the incorrect
Eq.~(\ref{Eq:theta-Lau}), respectively.
The red solid line represents the expected equilibrium distribution,
$P_\mathrm{eq}(\theta)=\sin\theta/2$.
The time scale is $\tau=R^2/(\mu k_\mathrm BT)$.
In both simulations, we use the Euler--Maruyama method with a time step $\Delta t/\tau=10^{-3}$,
a total of $n=10^5$ time steps, and $N=10^5$ particles.
}
\label{Fig}
\end{figure}

%%%%%%%%%%%%%%%%%%%%%%%%%%%%%%%%%%%%%%
\textit{Example I: A Brownian particle in a 2D harmonic potential}--
%%%%%%%%%%%%%%%%%%%%%%%%%%%%%%%%%%%%%%
Consider a Brownian particle in a harmonic potential
$\mathcal F=K_{ij}q^iq^j/2$ in a 2D flat space, where $K_{ij}$ is a positive-definite covariant tensor.
Note that, unlike the scalar free energy $\mathcal F$, the non-scalar free energy $F$
cannot be constructed as a coordinate-invariant quadratic form.
To determine $F$, the metric correction $\sqrt{\Delta}$ must be specified for a particular coordinate system,
as shown in Eq.~(\ref{FreeEnergy}).

For simplicity, we first introduce Cartesian coordinates $\mathbf q=(x,y)$ in the 2D flat space,
for which $K_{ij}=K\delta_{ij}$ ($i,j=x,y$), where $\delta_{ij}$ is the Kronecker delta and $K$ is a positive constant.
We assume that the particle experiences isotropic and homogeneous thermal noise, characterized by 
$D^{ij}=D\delta^{ij}$, where $D=k_\mathrm BT \mu$. 
The Christoffel symbols are $\Gamma_{jk}^i=0$.
Then, the scalar probability density in equilibrium is given by 
$\rho_\mathrm{eq}(x,y)=DK \exp [-K(x^2+y^2)/(2k_\mathrm BT)]/(2\pi k_\mathrm BT)$.

To illustrate how state-dependent diffusion emerges purely from a change of coordinates,
we now describe the same system in polar coordinates $\mathbf Q=(r,\phi)$,
defined by $(x,y)=(r\cos\phi,r\sin\phi)$.
The diffusion tensor transforms as
$D^{ab}=(\partial Q^a/\partial q^i)D^{ij}(\partial Q^b/\partial q^j)$,
where $a,b=r,\phi$.
The corresponding nonzero components of the diffusion metric are
$D_{rr}=1/D$ and $D_{\phi\phi}=r^2/D$, and the nonzero Christoffel symbols are
$\Gamma_{\phi\phi}^r=-r$ and
$\Gamma_{r\phi}^\phi=\Gamma_{\phi r}^\phi=1/r$.
Since $\Delta=(r/D)^2$, the equilibrium radial PDF becomes the Rayleigh distribution,
$P_\mathrm{eq}(r)=2\pi\sqrt{\Delta}\rho_\mathrm{eq}(r)
=Kr\exp[-Kr^2/(2k_\mathrm BT)]/(k_\mathrm BT)$.
Since the noise amplitudes satisfy $D^{ab}=B_{\alpha}^a B_{\alpha}^b$, they can be chosen as
$B_{1}^r=\sqrt{D}$, $B_{2}^\phi=\sqrt{D}/r$, and
$B_{2}^r=B_{1}^\phi=0$.

Using $\mathcal F(r)=Kr^2/2$ in Eq.~(\ref{Eq:LangevinA}), we obtain the C-LE for $r$,
\begin{align}
\dot r=-\mu Kr+\frac{D}{r}+\sqrt{2D}\xi_1,
\label{Eq:r}
\end{align}
which is independent of the stochastic convention parameter $C$.
Numerical integration of this LE yields the Rayleigh distribution shown in Fig.~\ref{Fig}(a).

For comparison, the corresponding non-scalar free energy obtained from Eq.~(\ref{FreeEnergy})
takes the form $F(x,y)=K(x^2+y^2)/2+k_{\rm B}T\ln D$ in Cartesian coordinates, whereas in polar coordinates 
it becomes $F(r,\phi)=Kr^2/2-k_{\rm B}T\ln(r/D)$.
These expressions clearly show that the non-scalar free energy is not coordinate invariant,
$F(x,y)\ne F(r,\phi)$.
Although the LL-LE~(\ref{Eq:LangevinLau}) yields the correct LE~(\ref{Eq:r}) when evaluated with $F(r,\phi)$,
incorrectly identifying $F(r,\phi)$ with $F(x,y)$ in Eq.~(\ref{Eq:LangevinLau})
leads to the erroneous equation:
\begin{align}
\dot r \overset{\mathrm{err}}{=} -\mu Kr+\sqrt{2D}\xi_1.
\label{Eq:r-Lau}
\end{align}
As shown in Fig.~\ref{Fig}(a), this incorrect LE results in a Gaussian distribution, which does not reproduce the invariant equilibrium measure in polar coordinates.
This example demonstrates that the geometric contribution encoded in the Christoffel term is essential 
even for diffusion in flat space written in curvilinear coordinates.

%%%%%%%%%%%%%%%%%%%%%%%%%%%%%%%%%%%%%%
\textit{Example II: A free Brownian particle on a spherical surface}--
%%%%%%%%%%%%%%%%%%%%%%%%%%%%%%%%%%%%%%
Consider a free Brownian particle constrained to move on a spherical surface of radius $R$.
The scalar free energy is therefore given by $\mathcal F=0$. 
Note again that, unlike the scalar free energy $\mathcal F$, the non-scalar free energy $F$
cannot be chosen to vanish in all coordinate systems.
The particle position is parameterized by $\mathbf q=(\theta,\phi)$,
where $\theta$ and $\phi$ are the polar and azimuthal angles, respectively.
The local isotropic diffusion coefficient in the tangent plane of the spherical surface is assumed to be $D$.
Due to the spherical geometry, the components of the diffusion metric depend on $\theta$:
$D_{\theta\theta}=R^2/D$, $D_{\phi\phi}=R^2\sin^2\theta/D$, and
$D_{\theta\phi}=D_{\phi\theta}=0$.
The nonzero Christoffel symbols are
$\Gamma_{\phi\phi}^\theta=-\sin\theta\cos\theta$ and
$\Gamma_{\phi\theta}^\phi=\Gamma_{\theta\phi}^\phi=\cot\theta$.
The noise amplitudes can be chosen as
$B_{1}^\theta=\sqrt{D}/R$ and
$B_{2}^\phi=\sqrt{D}/(R\sin\theta)$.

Using Eq.~(\ref{Eq:LangevinA}), we obtain the C-LE for $\theta$,
\begin{align}
\dot \theta=\frac{D}{R^2}\cot\theta +\frac{\sqrt{2D}}{R}\xi_1,
\label{Eq:theta}
\end{align}
which is independent of $C$.
Numerical integration of this LE yields the geometrically consistent equilibrium PDF
$P_{\rm eq}(\theta)=\sin\theta/2$, as shown in Fig.~\ref{Fig}(b).
This distribution correctly reflects the intrinsic surface measure on the sphere.

By contrast, the corresponding non-scalar free energy obtained from Eq.~(\ref{FreeEnergy}) is
$F(\theta,\phi)=-k_\mathrm BT\ln(R^2\sin\theta/D)$.
Although the LL-LE~(\ref{Eq:LangevinLau}) yields the correct LE~(\ref{Eq:theta}) when evaluated with $F(\theta,\phi)$,
incorrectly setting $F=0$ in the LL-LE~(\ref{Eq:LangevinLau}) leads to the erroneous equation:
\begin{align}
\dot \theta \overset{\mathrm{err}}{=} \frac{\sqrt{2D}}{R}\xi_1.
\label{Eq:theta-Lau}
\end{align}
As shown in Fig.~\ref{Fig}(b), this incorrect LE yields a uniform distribution, $P(\theta)=1/\pi$, which does not 
reflect the spherical geometry.

%%%%%%%%%%%%%%%%%%%%%%%%%%%%%%%%%%%
\textit{Example III: A Brownian particle on a curved surface}--
%%%%%%%%%%%%%%%%%%%%%%%%%%%%%%%%%%%
Consider a free Brownian particle confined to a curved surface embedded in 3D Cartesian space $(x,y,z)$.
In the Monge representation, the surface is described by
$\mathbf R(x,y)=(x,y,h(x,y))$, where $z=h(x,y)$ denotes its height~\cite{Naji2007,Ohta2020}.
We use the projected position $\mathbf q=(x,y)$ to parametrize the particle position on the surface, i.e.,
$\mathbf R=\mathbf R(\mathbf q)$.
Although the diffusion coefficient $D$ is homogeneous and isotropic in the local tangent plane,
its projection onto the $(x,y)$ plane gives rise to state-dependent diffusion.

The scalar free energy of the free particle is $\mathcal F=0$.
The diffusion metric and its determinant are
$D_{ij}=(1/D)[\delta_{ij}+(\partial_i h)(\partial_j h)]$
and
$\Delta=[1+(\partial_i h)^2]/D^2$, respectively, where $i,j=x,y$.
The corresponding diffusion tensor is
$D^{ij}=D\delta^{ij}-(\partial^i h)(\partial^j h)/(D\Delta)$,
where $\partial^i=\delta^{ik}\partial_k$.
The Christoffel symbols are
$\Gamma_{jk}^i=D^{il}(\partial_j\partial_k h)(\partial_l h)/D$,
and the noise amplitudes can be chosen as
$B_{\alpha}^i=\sqrt{D}[\delta_{\alpha}^i-(\partial_{\alpha}h)(\partial^i h)/(D^2\Delta+D\sqrt{\Delta})]$,
where
$\delta_1^x=\delta_2^y=1$, $\delta_1^y=\delta_2^x=0$, and
$\partial_{\alpha}=\delta_{\alpha}^j\partial_j$.

Using Eq.~(\ref{Eq:LangevinA}), the C-LE in the It\^o convention ($C=0$) becomes
\begin{align}
\dot q^i
=
-D^{jk}(\partial_j\partial_k h)
\frac{\partial^i h}{D^2\Delta}
+\sqrt{2}[B_{\alpha}^i\bullet\xi_\alpha]_0,
\label{NajiLE}
\end{align}
which coincides with the expression derived from geometric considerations by Naji and Brown~\cite{Naji2007}.
By contrast, if one incorrectly sets $F=0$ in the LL-LE~(\ref{Eq:LangevinLau}),
one obtains an erroneous equation.

%%%%%%%%%%%%%%%%%%%
\textit{Summary and discussion}--
%%%%%%%%%%%%%%%%%%%
In this work, we have formulated the C-LE~(\ref{Eq:LangevinA}) for stochastic dynamics with state-dependent diffusion.
The formulation is based on the scalar probability density $\rho$, the scalar free energy $\mathcal F$, and the diffusion metric defined as the inverse of the diffusion tensor $D^{ij}$.
Within this framework, the conventional spurious drift $\partial_jD^{ij}$ in the LL-LE~(\ref{Eq:LangevinLau}) is expressed as the Christoffel contribution $-\Gamma^i_{jk}D^{jk}$, as in Eq.~(\ref{Eq:LangevinA}).
The C-LE~(\ref{Eq:LangevinA}) is mathematically equivalent to the LL-LE~(\ref{Eq:LangevinLau}) through Eq.~(\ref{FreeEnergy}), but provides a geometrically transparent representation in terms of a scalar thermodynamic potential $\mathcal F$.
The covariance of the dynamics is particularly transparent in the Stratonovich form, Eq.~(\ref{Eq:LangevinS}), where the stochastic increment transforms as a contravariant vector.

The proposed C-LE~(\ref{Eq:LangevinA}) reproduces the correct equilibrium distributions in representative examples of state-dependent diffusion arising from coordinate transformations, geometric confinement, and projection from curved to flat space.
In 2D polar coordinates, it yields the Rayleigh distribution for the radial coordinate, while on a spherical surface it gives the equilibrium distribution proportional to $\sin\theta$.
For a Brownian particle on a curved surface in the Monge representation, the It\^o form of the C-LE agrees with the geometrically derived LE~\cite{Naji2007}.
These examples demonstrate that the Christoffel drift provides the geometric contribution required to preserve the canonical distribution associated with the configuration-space measure.

Although the C-LE~(\ref{Eq:LangevinA}) and LL-LE~(\ref{Eq:LangevinLau}) describe the same stochastic dynamics,
they differ in how the drift is decomposed into thermodynamic and geometric contributions.
The LL-LE incorporates the geometric contribution into the non-scalar restricted free energy $F$, whereas the C-LE separates the scalar free energy $\mathcal F$ from the geometric drift.
Our representation offers three practical advantages:
(i) The scalar free energy $\mathcal F$ can be constructed directly from standard symmetry arguments~\cite{ChaikinBook,VitelliBook}, whereas 
the restricted free energy $F$ cannot generally be constructed in this way in arbitrary coordinates.
(ii) The C-LE avoids the nontrivial log-Jacobian transformation of $F$ under coordinate changes.
Overlooking this contribution and incorrectly 
treating $F$ as a scalar free energy can lead to erroneous equations, such as Eqs.~(\ref{Eq:r-Lau}) and~(\ref{Eq:theta-Lau}).
(iii) The C-LE explicitly separates thermodynamic and geometric contributions to the drift, thereby making both their physical interpretation and the covariance of the dynamics more transparent.

Geometric formulations of stochastic dynamics have previously been discussed by Graham, Polettini, and others~\cite{Graham77,Graham1985,Morse2004Theory,Polettini2013}.
The novelty of the present work lies not in covariance itself, but in formulating state-dependent Langevin dynamics in terms of a scalar free energy and the diffusion metric.
This formulation identifies the conventional spurious drift as a Christoffel contribution and makes explicit the separation between thermodynamic and geometric contributions to the dynamics.

More broadly, the requirement that the free energy be a scalar is important not only for equilibrium distributions but also for path-level descriptions.
In covariant formulations of the Onsager and Onsager-Machlup principles, the Rayleighian, path probability, and entropy production are constructed from scalar thermodynamic quantities~\cite{yasuda2026covariant}.
The present C-LE~(\ref{Eq:LangevinA}) is consistent with this covariant variational framework at the level of stochastic dynamics.

The present formulation uses the Levi-Civita connection $\Gamma_{jk}^i$, characterized by metric compatibility, $\nabla_kD^{ij}=0$, and vanishing torsion, $\Gamma_{jk}^i=\Gamma_{kj}^i$.
These properties provide the natural geometric structure for the diffusion metric considered here.
More general configuration spaces involving odd diffusivity~\cite{Kalz24} or dislocations~\cite{Kobayashi24} may require extensions to connections with nonzero torsion, such as the Cartan connection~\cite{Yavari12}.
Extensions to inhomogeneous diffusion on curved manifolds, including spherical surfaces, also provide interesting directions for future work.

%%%%%%%%%%%%%%%%%
\textit{Acknowledgements}--
%%%%%%%%%%%%%%%%%
K.Y. is grateful to T. Uneyama for fruitful discussions and acknowledges JSPS KAKENHI for Grant-in-Aid for Early-Career Scientists (Grant No.\ 25K17357). 
K.I.\ acknowledges the Japan Science and Technology Agency (JST), FOREST (Grant No.\ JPMJFR212N), and CREST (Grant No.\ JPMJCR25Q1).
K.Y.\ and K.I.\ acknowledge the Japan Society for the Promotion of Science (JSPS) KAKENHI for Transformative Research Areas A (Grant No.\ 21H05309). 
S.K.\ acknowledges the support from the National Natural Science Foundation of China (Grant No.\ 12274098) and 
from the Zhejiang Key Laboratory of Soft Matter Biomedical Materials (2025ZY01036 and 2025E10072).
This work was supported by the JSPS Core-to-Core Program ``Advanced core-to-core network for the physics of self-organizing active matter" (JPJSCCA20230002).

%\bibliographystyle{sim} % 参考文献のスタイル（jplain, unsrtなど）
%\bibliography{references} % .bibファイル名を指定（拡張子は不要）

%%%%%%%%%%%%%%%%

%\appendix

\end{document}